\documentclass[useAMS,usenatbib]{mn2e}
\usepackage{float}
\usepackage{graphicx}
\usepackage{amssymb}
\usepackage{amsmath}
\usepackage{datetime}
\usepackage[normalem]{ulem}
\usepackage{times}
\usepackage[export]{adjustbox}

\usepackage{color}
\usepackage{soul}
\definecolor{stan}{rgb}{0,0,1}

\definecolor{steve}{rgb}{0,1,0}

\newcommand{\vinf}{V_\infty}
\newcommand{\vth}{v_{\rm th}}

\newcommand{\hturb}{H}

\newcommand{\beq}{\begin{equation}}
\newcommand{\eeq}{\end{equation}}
\newcommand{\beqa}{\begin{eqnarray}}
\newcommand{\eeqa}{\end{eqnarray}}

\title[Turbulent Porosity]{
Characterizing the turbulent porosity of stellar-wind structure generated by the line-deshadowing instability
}
\author[Owocki and  Sundqvist]{Stanley P.\ Owocki$^1$
and Jon O.\ Sundqvist$^2$
\\
$^1$Bartol Research Insitute, 
Department of Physics \& Astronomy, 
University of Delaware, Newark, DE 19716 USA
\\
$^2$KU Leuven, Instituut voor Sterrenkunde.
Celestijnenlaan 200D, 3001 Leuven, Belgium
}

\begin{document}




\maketitle

\label{firstpage}

\begin{abstract}
We analyze recent 2D simulations of the nonlinear evolution of the line-deshadowing instability (LDI) in hot-star winds, to quantify 
 how the  associated highly clumped density structure can lead to a ``turbulent porosity'' reduction in continuum absorption and/or scattering.
The basic method is to examine the statistical variations of mass column as a function of path length, and fit these to analytic forms that lead to simple statistical scalings for the associated mean extinction.
A key result is that one can characterize porosity effects on continuum transport in terms of a single ``turbulent porosity length", 
found here to scale as $\hturb \approx (f_{\rm cl} - 1) a$, where  $f_{\rm cl} \equiv \left < \rho^2 \right >/\left < \rho \right >^2$ is the clumping factor in density $\rho$,
and $a$ is the density autocorrelation length.
For continuum absorption or scattering in an optically thick layer, we find the associated effective reduction in opacity scales as $\sim 1/\sqrt{1+\tau_{\rm \hturb}}$,
where $\tau_{\rm \hturb} \equiv \kappa \rho \hturb$ is the local optical thickness of this porosity length.
For these LDI simulations, the inferred porosity lengths are small, only about a couple percent of the stellar radius, $\hturb  \approx 0.02 R_\ast$.
For continuum processes like bound-free absorption of X-rays that are only marginally optically thick throughout the full stellar wind,
 this implies $\tau_{\rm \hturb} \ll 1$, and thus that LDI-generated porosity should have little effect on X-ray transport in such winds.
 The formalism developed here could however be important for understanding the porous regulation of continuum-driven, super-Eddington outflows from luminous blue variables.
\end{abstract}


\section{Introduction}

Many astrophysical media exhibit extensive, complex, stochastic variations in density and/or opacity, often as a consequence of processes that generate supersonic, compressible turbulence.
A common example is the complex structure in the interstellar medium (ISM), 
wherein highly supersonic supernova explosions and stellar wind outflows compress ISM material into strong shocks, 
which through myriad interactions can develop into extensive, 3D stochastic structure characterized by strong variations in material density 
\citep[see, e.g.,][and references therein]{Squire17}.

A more specific example, which is the principal focus here,
regards the highly supersonic stellar winds of hot, luminous, massive stars
\citep[see review by][]{Puls08}, for which the driving by line scattering from bound-bound transitions of heavy, minor ions leads to a strong ``line-deshadowing instability" (LDI) 
\citep{MacGregor79,Carlberg80,Owocki84}.
Initial one-dimensional (1D) numerical radiation-hydrodynamical simulations of the nonlinear growth of this LDI showed the flow away from the wind base becomes highly structured, with high-speed  rarefied flow impacting and shock-compressing slower, denser shells
\citep{Owocki88,Feldmeier97b,Sundqvist15}.
In  multi-dimensional  simulations, these shells break up, by Rayleigh-Taylor instabilities and under the influence of oblique radiation rays, into a complex web of inter-connected dense clumps and partially refilled rarefied regions
\citep{Dessart03,Dessart05,Sundqvist17}.

A longstanding challenge has been to quantity the effect of this extensive wind clumping on observational diagnostics.
For example,  two-body processes like recombination or collisional excitation  are generally expected to be enhanced in proportion to a ``clumping factor'', defined by the ratio of the mean-squared-density to the square of the mean-density, $f_{\rm cl} \equiv \left < \rho^2 \right >/\left < \rho \right >^2$.
Among other things, this can lead to overestimation of observationally inferred mass-loss-rates from such stellar winds.
(See, e.g., reviews in \citet{Hamman08}.)

However, even for extinction processes that scale linearly with density, e.g. electron scattering or bound-free absorption, the effective opacity can be {\em reduced} by the medium's ``porosity'', associated with the enhanced radiation transport through relatively low-density regions
\citep{Shaviv98,Owocki04,Oskinova04,Oskinova07,Sundqvist14}.

For example, in the idealized case of a medium in which {\em all} material is confined into spatially isolated, dense clumps with characteristic size $l_{\rm cl}$ and mean separation $L$, we can define a ``porosity length'' $h$ in terms of the mean-free-path between clumps, given by the ratio of the empty volume to the clump cross section,  $(L^3-l_{\rm cl}^3)/l_{\rm cl}^2$. Since the volume filling fraction $f_{\rm vol}=(l_{\rm cl}/L)^3 = 1/f_{\rm cl}$, this porosity length scales simply as 
\beq
h = (f_{\rm cl}-1) l_{\rm cl}
\, .
\label{eq:hfclcl}
\eeq
For a microscopic absorption opacity $\kappa$,  the associated optical thickness $\tau_h = \left < \kappa \rho \right > h$ of this porosity length sets the effective reduction in opacity
\citep[see, e.g., eqn.\ (35) of][]{Owocki04},
\beq
\frac{\kappa_{\rm eff}}{\kappa} = \frac{1 - e^{-\tau_h}}{\tau_h} 
\, .
\label{eq:kapeffbkap}
\eeq
For $\tau_h \ll 1$, the effective opacity recovers this microscopic value;  
in particular, in the limit of smooth, unclumped medium with $f_{\rm cl} \rightarrow 1$, we have $h \rightarrow 0$ and so $\tau_h \rightarrow 0$, implying $\kappa_{\rm eff} \rightarrow \kappa$.
In contrast, for $\tau_{\rm h} \gg 1$, the effective opacity is {\em reduced} by a factor $1/\tau_h$
\citep{Shaviv98,Owocki04,Oskinova04}.

Various extensions have been proposed, e.g.\ to account for a non-vanishing density in the interclump medium \citep{Sundqvist14},
or a distribution of clump sizes \citep{Owocki04}, and these lead to somewhat different scalings for the reduction in opacity as function of $\tau_h$; 
but such previous parameterizations have generally still been grounded in this basic two-component paradigm consisting of distinct clumps embedded in a background interclump medium.
As such, it has remained unclear how the concepts from such idealized two-component clumping models can be translated to characterize the porosity for actual dynamical simulations of the density structure in a medium with compressible turbulence.

The central goal of this paper is to develop a formalism for characterizing such 
``turbulent\footnote{The usage of ``turbulent'' here simply refers to a medium for which density fluctuations are effectively stochastic, with statistical properties that can be identified by a suitable averaging procedure;  it is {\em not} necessarily tied to any specific model or mathematical description for the turbulence, which in the compressible case can depend on details of the excitation mechanism(s).} porosity''.
To provide a specific, concrete example, we focus on recent results of 2D radiation hydrodynamical simulations of the extensive structure that arises from the LDI of a line-driven stellar wind
\citep{Sundqvist17}.
But the general formalism developed here could in principal be applied to characterize the porosity of any medium for which multi-dimensional hydrodynamical simulations show extensive stochastic variations in density, including, e.g., for the ISM.

As detailed in \S 2, the basic approach centers on characterizing the statistical distribution of column masses as a function of the length of  a  random path-integral through the medium.
Using this to define an {\em ensemble-average extinction}, we derive an analytic generalization of the usual exponential attenuation with optical depth $\tau$ that now also depends on the optical thickness $\tau_{\rm \hturb}$ of a numerically derived {\em turbulent} porosity length\footnote{We choose upper case $H$ notation to distinguish this turbulent porosity length from the conceptually related, but distinct form for the porosity length $h$ of an idealized two-component clump model.}$ \hturb$.
A key result (\S 2.5) is that this follows closely the scaling form (\ref{eq:hfclcl}) if we simply replace the clump size $l_{\rm cl}$ with the density autocorrelation length $a$, i.e.,  $H \approx (f_{\rm cl}-1)a$.
Relative to a smooth, unclumped medium, we show (\S 3.1) that the associated mean-free-path is enhanced by a factor that scales approximately with $\sqrt{1+\tau_{\rm \hturb}}$.  
Monte Carlo simulations of radiative transport (\S 3.2) through a planar scattering layer of fixed optical thickness $\tau$ shows that the usual scaling for transmission is modified from the smooth case by replacing $\tau$ with the ratio $\tau/\sqrt{1+\tau_{\rm \hturb}}$, thus implying an effective porosity reduction of the opacity by the factor $1/\sqrt{1+\tau_{\rm \hturb}}$.
For OB winds with continuum processes that have optical depth of order unity or less throughout the entire wind, this implies $\tau_{\rm \hturb} \ll 1$ and thus negligible porosity effects on continuum transport, like the bound-free absorption of X-rays (\S 4.1).
But for optically thick winds from Wolf-Rayet stars, or from luminous blue variable (LBV) eruptions, porosity effects could be significant if the flow develops extensive density structure, possibly affecting even the dynamics of the  driving or initiation of such outflows 
\citep{Owocki04} (\S 4.2).
While the detailed scaling results derived here are specific to this model of LDI-generated turbulence in stellar winds,  we conclude with some brief remarks on the potential broader application  of the basic formalism for characterizing porosity effects on the radiation transport in the ISM or other media with stochastic density structure associated with compressible turbulence.

\section{Extinction in a Turbulent Medium}

\subsection{Column mass probability distribution}

To characterize the porosity of a clumpy medium, let us first consider the optical depth through some random length segment $\ell$,
\beq
t (\ell) \equiv \int_0^\ell \kappa \rho \, d\ell
\, ,
\label{eq:taul}
\eeq
where $\rho$ and  $\kappa$ are the mass density and the opacity\footnote{The simplest case would be for a wavelength-independent (grey) opacity, such as from electron scattering. But a continuum wavelength dependence, such as from bound-free absorption, could be also accounted for by treating multiple wavelength bins independently. See \S \ref{sec:weakpor}.} 
(a.k.a. mass-absorption coefficient) along the segment, 
which is assumed to be short compared to any background gradient scale for the mean (ensemble-averaged) density or opacity. 
For a turbulent medium in which these quantities exhibit some stochastic variations, let us define an ensemble-averaged optical depth,
\beq
\tau(\ell) \equiv \left < t (\ell) \right >
\, ,
\label{eq:tdef}
\eeq
along with an associated measure,
\beq
m (\ell ) \equiv \frac{t (\ell )}{\tau (\ell )}
\, ,
\label{eq:mdef}
\eeq
which characterizes the random variation of this optical depth relative to this ensemble-average.
While this formulation is general to include cases with a variable (e.g., density--dependent) $\kappa$, note that
for the specific case of a fixed (density-{\em in}dependent) $\kappa$, the quantity $m$ just represents 
the variation of the {\em mass column density} relative to its mean;
for convenience, we shall thus refer to $m$ as the ``normalized mass column''.

With these definitions, we can next define the associated ensemble-averaged extinction as
\beq
E(\tau) \equiv \left < e^{-m \tau} \right > \equiv \int_0^\infty \frac{df}{dm} \, e^{-m \tau} \, dm
\, ,
\label{eq:Edef}
\eeq
wherein the latter equality defines the fractional distribution of mass column, $df/dm$.

In a smooth medium, this distribution is simply given by a delta-function, $df/dm = \delta (m-1)$, so that the extinction defined in eqn.\ (\ref{eq:Edef}) reduces to its usual exponential form $E(\tau) = e^{-\tau}$.

But for a medium with extensive density and/or opacity structure, we see from (\ref{eq:Edef}) that the generally nonlinear, exponential weighting over optical depth will tend to favor less opaque paths, leading to a potential porosity reduction in the average extinction.
Within this formalism, characterizing this porosity thus essentially requires determining this mass column distribution $df/dm$ and its variation with the path length $\ell$.

\subsection{Numerical results from LDI simulations}

As a specific example, we apply here results from our recent 2D simulations of the strong  line-deshadowing instability  (LDI) of radiatively driven stellar winds
\citep{Sundqvist17}.
Within a pseudo-planar geometry extending vertically by one stellar radius $R_\ast$ from the stellar surface, these 2D, isothermal wind simulations account for lateral variation in structure over a horizontal period set to $\Delta X = R_\ast/10$.
The associated spatial grid is uniform, with 1000 radial zones and 100 lateral zones.

\begin{figure}
\begin{center}
\includegraphics[scale=1.25]{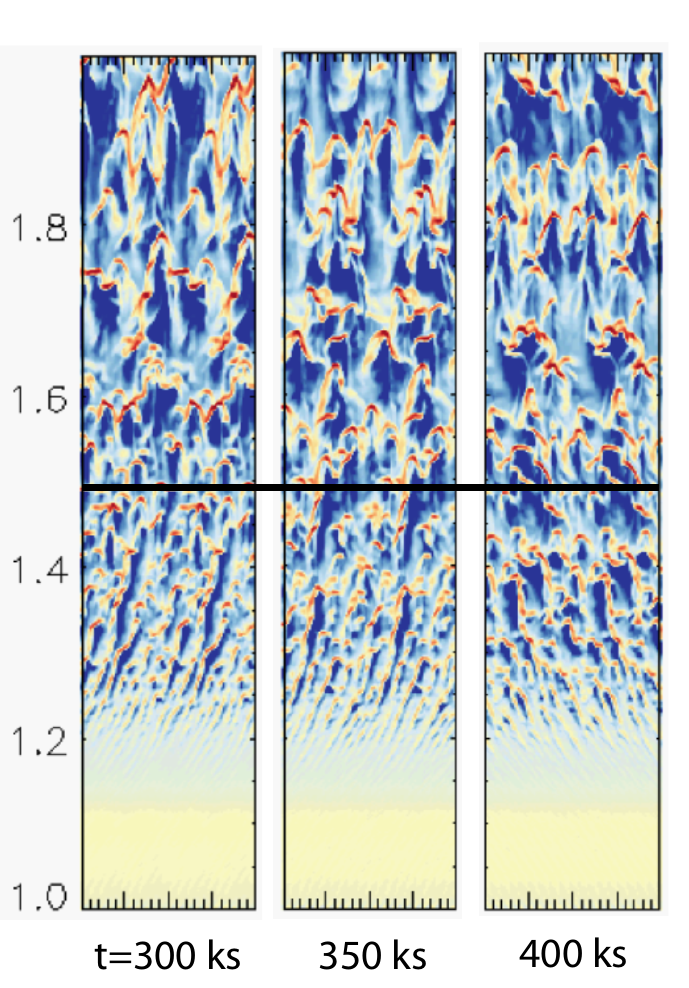}
\caption{For a 2D pseudo-planar radiation-hydrodynamical simulation of instability-generated structure in a line-driven stellar wind, spatial and temporal variation of log density, relative to the initial CAK steady-state, with  color scales ranging from $\log (0.1)=-1$ for lowest density (blue) to $\log(10)=+1$ for highest density (red),  with yellow ($\log(1)=0$) representing densities equal to the initial, CAK steady-state. The vertical variation extends from the stable, subsonic wind base at the stellar surface, to a height of one stellar radius $R_\ast$. The lateral variation is over {\em twice} the horizontal periodic box length of $\Delta X = 0.1 \, R_\ast$. The simulation assumes uniform grid spacing of  $dx = dz = 0.001 \, R_\ast$ in both the lateral (x) and vertical (z) directions.  
The 3 panels show time snapshots at $t=$300, 350, and 400\,ks, long after the initial CAK condition has developed into a statistically steady turbulent state. 
The statistical quantities computed here are taken from the region above the horizontal thick line at $r=1.5 R_\ast$, at times $t>300$\,ks after the structure in this region
has fully developed into a stochastically varying, quasi-steady state.}
\label{fig:fig1}
\end{center}
\end{figure}

Figure \ref{fig:fig1} shows the radial and horizontal variation in density relative to the initial steady-state, set by the steady-state, line-driven wind model of 
\citet[][hereafter ``CAK'']{Castor75}, for  time snapshots $t=$300, 350, and 400\,ks, long after adjustment to a turbulent state, 
The statistical results presented below are for times $t>300$\, ks and from the upper half of the simulation, where the LDI-generated structure has fully developed.

\begin{figure}
\begin{center}
\includegraphics[scale=.65]{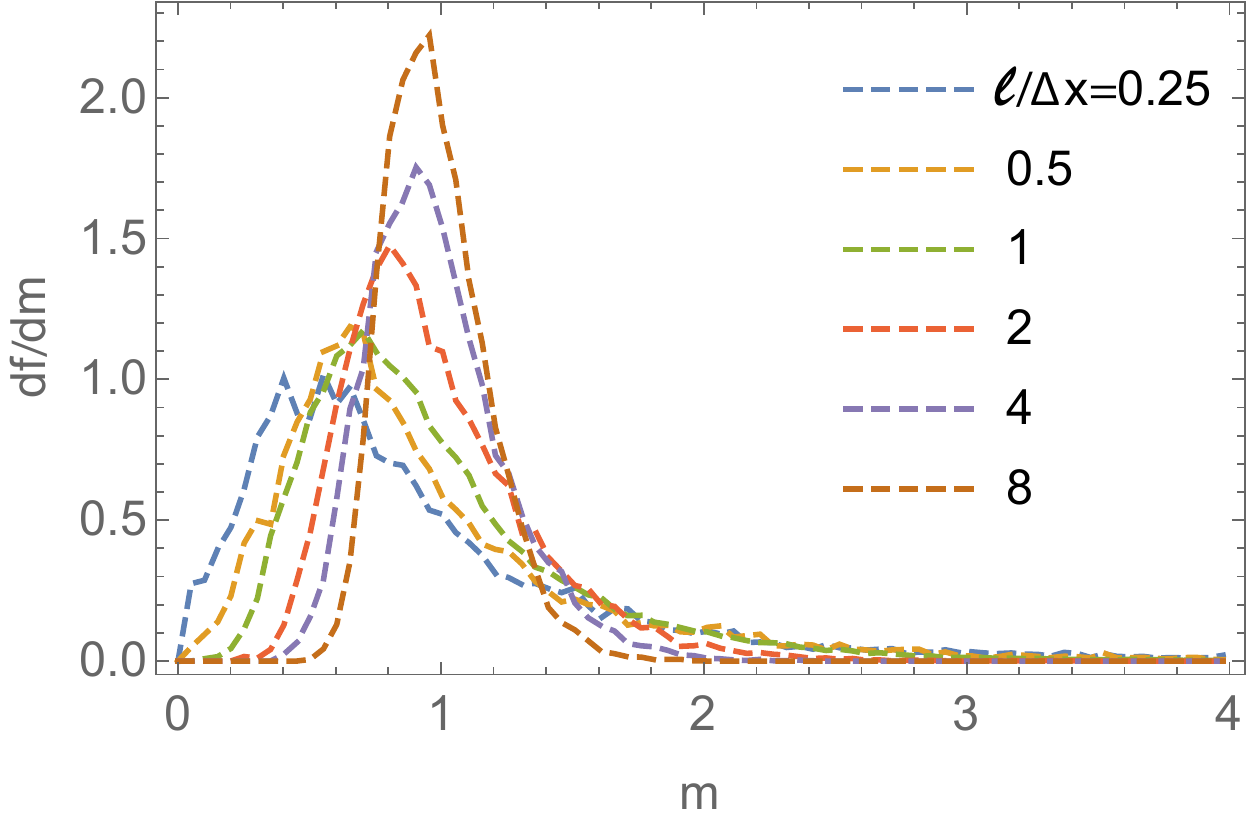}
\includegraphics[scale=.65]{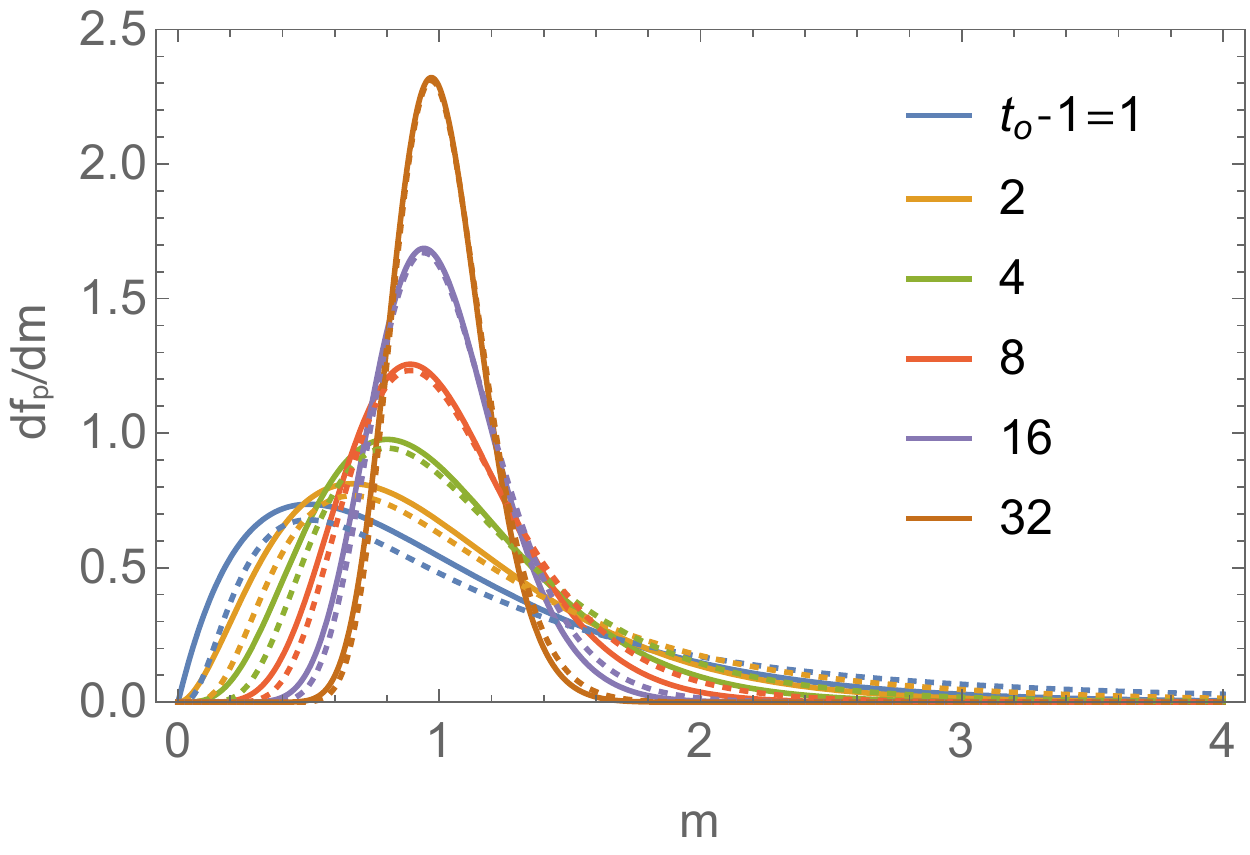}
\caption{
Mass column distribution functions $df/dm$ plotted vs.\ normalize mass column $m$.
 The top panel (a) shows empirical results from LDI simulations for the labeled path lengths $\ell$.
The lower panel (b) shows corresponding analytic fitting functions for these empirical distributions.
The solid curves show the exponentially truncated power law of eqn.\ (\ref{eq:dfdmc2}) for the labeled values of $t_{\rm o}  = 1 + \ell/\hturb$,
representing the corresponding fits to empirical LDI distributions for various path lengths $\ell$, with the turbulent porosity length $\hturb = 0.25 \, \Delta X$.
The dotted curves compare log-normal distribution functions that peak at the same mass column, implying widths that scale as $\sigma =\sqrt{\log(t_{\rm o}/(t_{\rm o}-1))}$.
}
\label{fig:fig2}
\end{center}
\end{figure}

For a very large number ($>$50,000) of random samples through the clumped wind structure, figure \ref{fig:fig2}a plots the resulting numerical distributions $df/dm$ vs.\ $m$ for path lengths\footnote{For  lengths greater than a single horizontal period, $\ell > \Delta X$, the renditions of neighboring periods are randomized in phase to preserve the stochastic nature of the horizontally extended structure.} $\ell/\Delta X$=0.25, 0.5, 1, 2, 4, and 8.
Clearly the form and width of the distribution depends sensitively on the path length.
Smaller $\ell$ give a broad distribution, reflecting the wide range in local mass density.
But for large $\ell$, these density variations are averaged out, leading to a narrowing of the distribution function, which in the limit of very large $\ell$ approaches in effect the delta-function form that would apply in a completely smooth medium.

\begin{figure}
\begin{center}
\includegraphics[scale=.46]{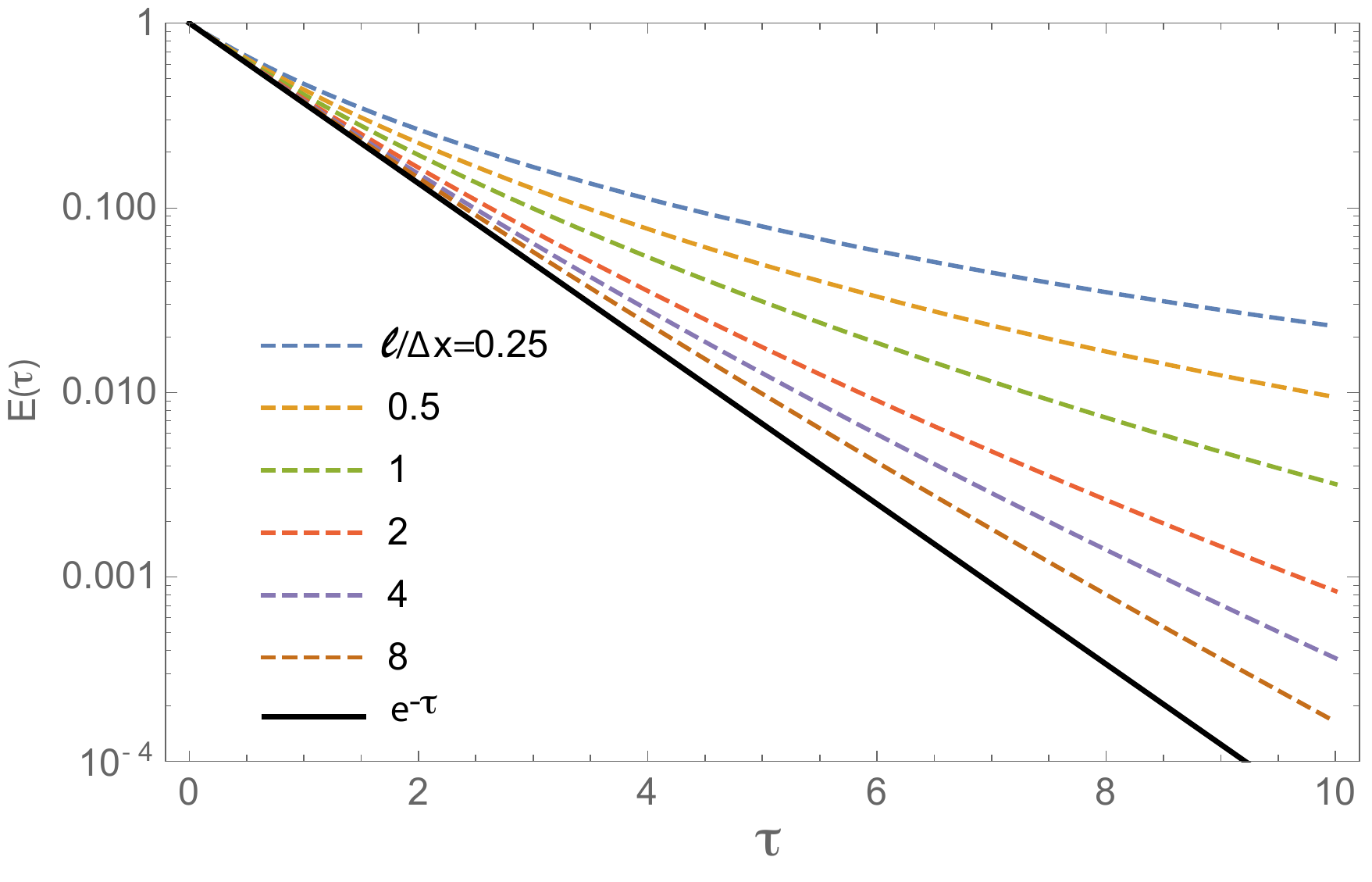}
\,
\includegraphics[scale=.46]{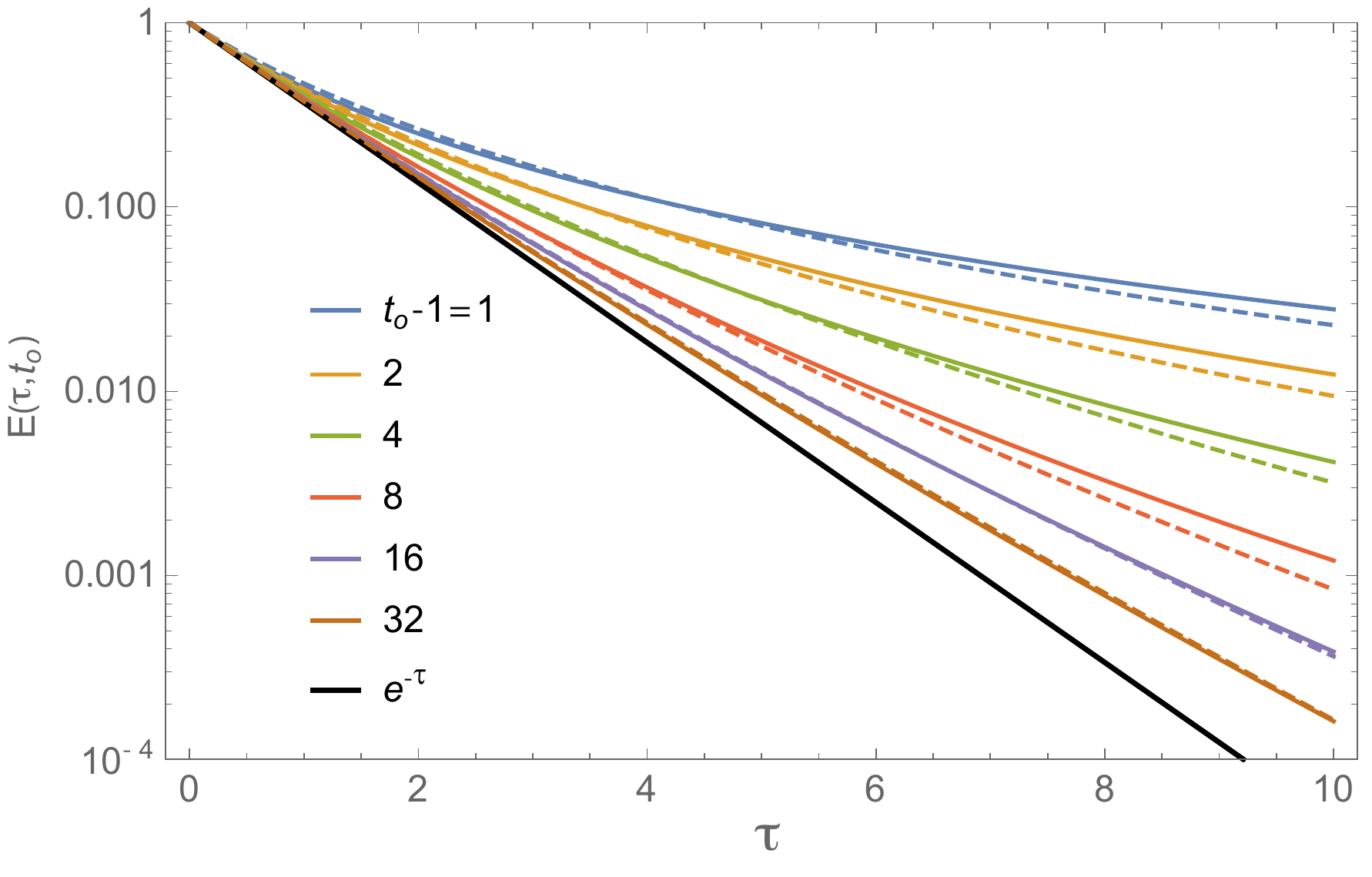}
\caption{
{\em Top:} Numerical results for extinction vs. optical depth, $E(\tau)$, as computed from eqn.\ (\ref{eq:Edef}) from numerical column-mass distributions plotted in figure \ref{fig:fig2}a for the labeled path lengths $\ell$. 
{\em Bottom:}
The dashed curves again reproduce the top panel, while the solid curves now compare these with analytic extinction function $E(\tau ; t_{\rm o})$ derived in eqn.\ (\ref{eq:Etto}) from the truncated power-law fit (\ref{eq:dfdmc2}) to the mass distribution, for the labeled values of the power exponent $t_{\rm o}$.
The close agreement indicates a close relationship between path length $\ell$ and power exponent, viz.\ $t_{\rm o} \approx 1 + 4 \ell/\Delta X$.
}
\label{fig:fig3}
\end{center}
\end{figure}

These differences translate into an associated path-length dependency of porosity effects.
This can be readily seen by applying these numerical distributions within the mass-column sum in eqn.\ (\ref{eq:Edef}) to compute the ensemble-averaged extinction.
For the same sample of path lengths shown in figure \ref{fig:fig2}a, the labeled dashed curves in figure \ref{fig:fig3}a plot the resulting extinction as function of mean optical depth, $E(\tau)$, as computed from applying the numerical results for $df/dm$ into eqn.\ (\ref{eq:Edef}).

Comparison with the solid black line -- representing in this semi-log plot a purely exponential extinction -- immediately shows how the porosity reduction in extinction depends on the sample length.
This porosity reduction is strongest for the shortest path lengths, but the effect is weakened by averaging over longer lengths; indeed, in the large-length limit the extinction asymptotically approaches the exponential form of a smooth medium.
 
\subsection{Exponentially truncated power-law fit for LDI column-mass distribution}

While these numerical results provide a basis for characterizing the porosity of this turbulent medium, establishing their overall scaling in terms of porosity parameters requires further analysis grounded in parameterized fitting functions for the numerical distribution $df/dm$.

One immediate possibility is the log-normal form often found to characterize statistical distributions.
For the case here of normalized distribution with unit mean $\left < m \right >=1$ (and thus $\ln \left < m \right > = 0$), this takes the form
\beq
\frac{df_{\rm l}}{dm} = \frac{e^{-(\ln(m)/\sigma)^2}}{m \sigma \sqrt{2 \pi}}
\, ,
\label{eq:lognorm}
\eeq
where the single parameter $\sigma$ characterizes the width of the distribution.
We do indeed find that this log-normal distribution can provide a potentially viable fit (see the dotted curves in figure \ref{fig:fig2}b), but its application to integral averaging for extinction in eqn.\ (\ref{eq:Edef}) does not yield a suitable analytic form.

Instead, let us pursue fitting the numerical distributions with an exponentially truncated power-law, 
\beq
\frac{df_{\rm p}}{dm} = C m^n e^{-m t_{\rm o}}
\, ,
\label{eq:dfdmc1}
\eeq
where $C$ is a normalization constant that is set by requiring the integrated fraction over all $m$ is unity. 
Moreover, to ensure also that $\left < m  \right >=1$, we require that the ``porosity exponent" $t_{\rm o}= n+1$,
leading to a normalized distribution of the form,
\beq
\frac{df_{\rm p}}{dm} =   (m t_{\rm o})^{t_{\rm o}}  \frac{ e^{- m t_{\rm o}}}{ m \Gamma (t_{\rm o} )} ~~   ;  ~ t_{\rm o} > 1
\, .
\label{eq:dfdmc2}
\eeq
The solid curves of figure \ref{fig:fig2}b show this distribution for various values of the parameter $t_{\rm o}$.
As shown below (see eqn.\ \ref{eq:Etto}), application of this power-law form in the extinction integral (\ref{eq:Edef}) yields a simple analytic scaling for the extinction.

The restriction $t_{\rm o} > 1$ ensures $df/dm  = 0$ for $m=0$, as appropriate for a physically realistic medium with non-zero density everywhere.
Order-unity $t_{\rm o}$ imply strong porosity, while larger $t_{\rm o}$ imply weaker porosity.
Indeed, in the limit $t_{\rm o} \rightarrow \infty$, the distribution approaches a delta-function,
\beq
\frac{df_{\rm p}}{dm}= \delta (m-1)  ~;~   t_{\rm o} \rightarrow \infty
\, .
\label{eq:dfdmdel}
\eeq
Since all the mass columns then just have the mean value, this represents a smooth medium, without any clumping or porosity;
its application in eqn.\ (\ref{eq:Edef}) thus simply recovers the usual exponential form for extinction.

\subsection{Turbulent porosity length $H$ and associated porosity thickness $\tau_H$}

By connecting this parameterized function to associated numerical results, we can now use the inferred dependence of the parameter $t_{\rm o}$ on path length $\ell$ to parameterize the turbulent porosity in terms of a characteristic length $H$ and its associated optical depth $\tau_H$.

Figure \ref{fig:fig2}b plots $df/dm$ vs.\ $m$ for various $t_{\rm o}$, selected to approximately fit the numerical distributions in figure \ref{fig:fig2}a.
Much as with the numerical distribution's trend from broad to narrow with increasing path length, these power-law distributions become narrower with increasing $t_{\rm o}$, indicating a direct link of the exponent $t_{\rm o}$ with path length $\ell$ (see eqn. (\ref{eq:toell})).
The solid curves show the exponentially truncated power-law function given by eqn.\ (\ref{eq:dfdmc2}), while the dotted curves 
show corresponding log-normal distributions of eqn.\ (\ref{eq:lognorm}), with widths set to $\sigma = \sqrt{\ln(t_{\rm o}/(t_{\rm o}-1)}$, 
which ensures a distribution peak at the same $m$ as the truncated power-law.
The close overlap shows that both analytic functions could represent comparably good fits to the numerical distributions.

However, the power-law distribution has the advantage of yielding an analytic form for the ensemble-averaged extinction,
\beq
E(\tau; t_{\rm o})
= \int_0^\infty (m t_{\rm o})^{t_{\rm o}}  \frac{ e^{- m (\tau+ t_{\rm o})}}{ m \Gamma (t_{\rm o} )}  \, dm 
= \left ( 1 + \frac{\tau}{t_{\rm o}} \right )^{-t_{\rm o}}
\, .
\label{eq:Etto}
\eeq
The last expression represents an important initial result of our analysis. It shows that, in a porous medium with $t_{\rm o} \gtrsim 1$, extinction over a fixed path length follows a {\em power-law} scaling.
However, for distributions with $t_{\rm o} \gg 1$, straightforward Taylor expansion shows that this porous extinction simply recovers the usual exponential attenuation of a smooth medium,
\beqa
&& 
E(\tau; t_{\rm o}) =
\left ( 1+ \frac{\tau}{t_{\rm o}}  \right )^{ -t_{\rm o}} 
\nonumber
\\
&\approx& 
1 - \tau 
+ \frac{ \left ( 1+t_{\rm o} \right ) \tau^2}{2 t_{\rm o}} 
- \frac{ \left ( 2 + 3 t_{\rm o} + t_{\rm o}^2 \right ) \tau^3}{6 t_{\rm o}^2}
+ O[\tau]^4
\nonumber
\\
&\approx&
1 - \tau 
+ \frac{\tau^2}{2} 
- \frac{\tau^3}{6}
+ O[\tau]^4 ~ ; ~ t_{\rm o} \gg 1
\nonumber
\\
&\approx& e^{-\tau}
\, .
\label{eq:Ettoexp}
\eeqa

The upshot then is that there is a direct relationship between the porosity exponent $t_{\rm o}$ and the dependence of the width of the mass-column distribution $df/dm$ on the path length $\ell$.
The solid curves in figure \ref{fig:fig3}b plot the optical-depth variation of this power-law extinction for the same set of $t_{\rm o}$ used in figure \ref{fig:fig2}b.
The dashed curves overplot the corresponding results from the numerical distributions for the various lengths $\ell$.
The close agreement indicate the relationship between $t_{\rm o}$ and $\ell$ can be cast in the linear form,
\beq
\boxed{
t_{\rm o} \approx 1+ \frac{\ell}{\hturb} = 1+ \frac{\tau}{\tau_{\rm \hturb}}
}
\, ,
\label{eq:toell}
\eeq
where we find {\em empirically} that, for this LDI-generated structure, $\hturb \approx \Delta X/4$.
This now represents a new kind of characteristic scale for this turbulent porosity, which we thus refer to as the ``turbulent porosity length".

The latter equality in (\ref{eq:toell}) recasts this scaling in terms of the ratio of  the mean optical depth $\tau \equiv \left < \kappa \rho \right > \ell$ over the path length $\ell$ to a ``porosity optical thickness'', $\tau_{\rm \hturb} \equiv  \left < \kappa \rho \right > \hturb $.
Within this formalism, this turbulent porosity length $\hturb$ simply corresponds to the length scale over which the porosity exponent $t_{\rm o}=2$, yielding a power-law scaling for the extinction,
$
E(\tau; 2) = 1/ \left ( 1 + \tau/2 \right )^{2}
$.

More generally, applying (\ref{eq:toell}) in (\ref{eq:Etto}), and so defining $E(\tau,\tau_{\rm \hturb} ) \equiv E(\tau; t_{\rm o}=1+\tau/\tau_{\rm \hturb} )$, 
we now obtain a scaling for the extinction with mean optical depth $\tau$ in a medium with turbulent porosity optical thickness $\tau_{\rm \hturb}$,
 \beq
\boxed{
E(\tau,\tau_{\rm \hturb})
 = \left (1 + \frac{\tau}{1+ \tau/\tau_{\rm \hturb}} \right )^{-(1+\tau/\tau_{\rm \hturb})}
  }
\, 
\label{eq;Etth}
\eeq
The close agreement between the analytic (solid curves) and numerical (dashed curves) results in figure \ref{fig:fig3}b provide strong evidence for the analytic formulation for porous extinction given by eqn.\ (\ref{eq;Etth}).
In conjunction with the identification of the turbulent porosity length $\hturb$ and its associated optical thickness $\tau_{\rm \hturb}$, this thus represents a key result for characterizing the porosity effects on radiation transport in this medium with LDI-generated clumping.

\subsection{Porosity in terms of autocorrelation length and clumping factor}

Let us next seek a simpler, intuitive basis for the physical origin of the value of turbulent porosity length $H = \Delta X/4 = 0.025 R_\ast$,
specifically in terms of a density autocorrelation length $a$ and the associated density clumping factor $f_{\rm cl}$.

 Linear stability analyses \citep{Owocki84} show that the LDI applies at scales near and below the ``Sobolev length", $L_{\rm Sob} \equiv \vth/(dv/dr)$ \citep{Sobolev60}, where $\vth$ is the ion thermal speed\footnote{This ignores broadening from any `microturblent' velocities, e.g. associated with any variations or waves from the underlying star, and/or from the LDI itself.}; for typical values $\vth \approx 10$\,km\,$s^{-1}$ and $dv/dr \approx \vinf/R_\ast$ with wind terminal speed $\vinf \approx 2000$\,km\,$s^{-1}$, this gives $L_{\rm Sob} \approx 0.005 \, R_\ast$.
The recent 2D simulations of the nonlinear evolution of this LDI find  that the resulting structure in the outer wind
has a density autocorrelation length\footnote{\citet{Sundqvist17} quote the autocorrelation as FWHM $\approx 0.01 R_\ast$, referring to the full-width at half-maximum of the autocorrelation function. The quantity $a$ here refers to the $\sigma$ value of the best-fit gaussian, with $a \approx $ FWHM/2.}  comparable to this Sobolev length, $a \approx 0.006 R_\ast \approx L_{\rm Sob}$ \citep[see the lower left panel of figure 6 from][]{Sundqvist17}.
In these simulations, this scale is set physically by the lateral components of the line-force. This stands in marked contrast to results previous instability models that do not account for such lateral radiation forces \citep{Dessart03}, which show lateral structure extending down to the grid scale.
But since the autocorrelation length $a$ here is resolved over several ($\sim 6$) grid points of size $0.001 R_\ast$, and yet is well below the horizontal period $\Delta X$, it is likely to be insensitive to these specific numerical choices for grid resolutions and horizontal period.

The upper left panel of figure 6 from \citet{Sundqvist17}
plots the radial variation of the density clumping factor $f_{\rm cl} \equiv \left < \rho^2 \right >/\left < \rho \right >^2$ for the saturated wind structure at evolved times $t>250$\,ks of these 2D LDI simulations. Note that for radii $r>1.5 R_\ast$, the clumping factor approaches a value $f_{\rm cl} \approx 4$.

Following the scaling (\ref{eq:hfclcl}) for the porosity of simple clump model, if we identify autocorrelation length $a$ with the clump size, $l_{\rm cl} \approx a \approx 0.006 R_\ast = 0.06 \, \Delta X$, we can use the derived clumping factor to estimate the associated porosity length 
\beq
H_{\rm cl} = (f_{\rm cl} -1) a \approx 0.18 \Delta X
\, .
\label{eq:Hcl}
\eeq
This is comparable to, but just slightly smaller than, the value $H= 0.25 \Delta X$ inferred empirically from the above, much more extensive $df/dm$ analysis. 

This indicates that the porosity from a dynamical simulation of a medium with compressible turbulence can be estimated by simply computing the density autocorrelation length $a$ and the clumping factor $f_{\rm cl}$, and then applying eqn.\ (\ref{eq:Hcl}) to compute the associated turbulent porosity length $H$.

\section{Porosity Modification of  Radiation Transport}

The above analysis provides a scaling to quantify porosity effects on the level of extinction associated with {\em absorption} of radiation.
More generally radiation interacting with matter can be subsequently {\em scattered} or {\em thermally reemitted}, leading in optically thick media to a diffusive spatial transport of the radiative energy.
For the simple case of an effectively gray, continuum opacity (including cases with an appropriate frequency-average such as the Rosseland mean), we can apply the above extinction model to characterize porosity effects on such spatial transport.
The Monte-Carlo scattering model in \S \ref{sec:mcscat} below provides an illustrative example, but as background for interpreting those results, let us first consider how porosity effects alter the average mean-path lengths for such diffusive radiation transport.

\subsection{Porosity increase in mean path length}

In an smooth medium, the mean optical-depth traversed by a photon is simply given in terms of a weighted average over  the exponentially declining probability to travel an optical depth $\tau$,
\beq
{\bar \tau} = 
{\int_0^\infty \tau \, e^{-\tau} \, d\tau}
= 1
\, .
\eeq
By analogy, let us define a normalized transport probability to traverse a mean optical depth $\tau$ in a medium with porosity depth $\tau_{\rm \hturb}$,
\beq
{\hat E} (\tau,\tau_{\rm \hturb} ) \equiv
\frac{ E(\tau,\tau_{\rm \hturb})}
{\int_0^\infty  \, E(t,\tau_{\rm \hturb}) \, dt} 
\, .
\label{eq:Ehatdef}
\eeq
The associated mean optical path in this porous medium can then be computed from
\beq
{\bar \tau} (\tau_{\rm \hturb} ) \equiv 
\int_0^\infty \tau \, {\hat E}(\tau,\tau_{\rm \hturb}) \, d\tau
\, .
\eeq
\begin{figure}
\begin{center}
\includegraphics[scale=0.55]{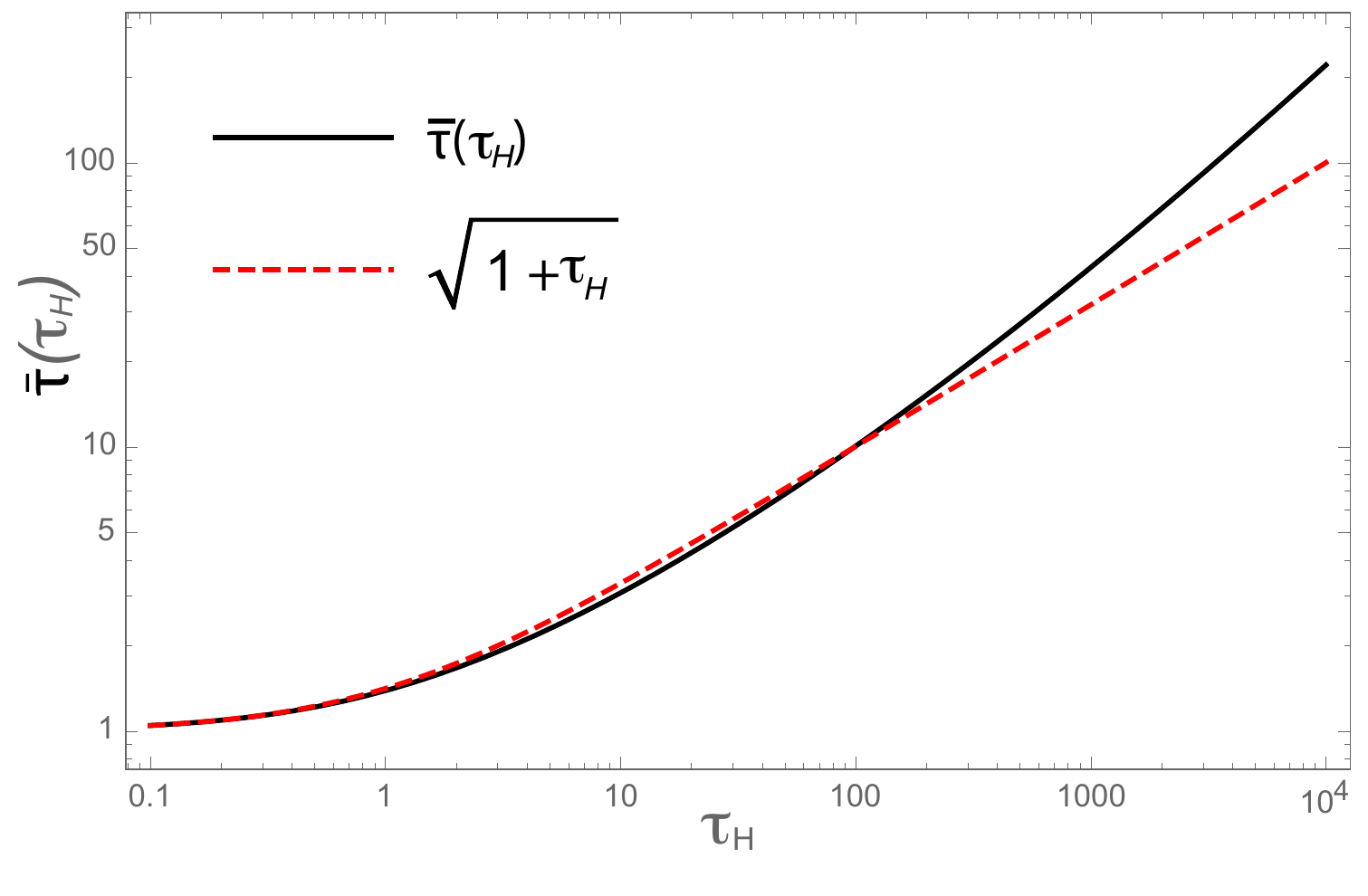}
\caption{
Mean-free path in a porous medium with extinction set by the analytic form eqn.\ (\ref{eq;Etth}), measured in terms of the mean optical depth $\bar{\tau}$, plotted vs. the porosity depth $\tau_h$.
The dashed red line shows this scaling can be approximately fit by the simple analytic scaling $\sqrt{1+\tau_{\rm \hturb}}$.
}
\label{fig:fig4}
\end{center}
\end{figure}
The solid black curve in figure \ref{fig:fig4} plots this  ${\bar \tau}$ vs. $\tau_{\rm \hturb}$; the dashed red curve shows that the variation is well fit by the simple scaling $\sqrt{1+\tau_{\rm \hturb}}$. 

The overall upshot here is that the turbulent porosity increase in the mean-path scales approximately with $\sqrt{1+\tau_{\rm \hturb}}$.

\subsection{Monte Carlo results for porous transport through a planar scattering layer}
\label{sec:mcscat}

\begin{figure}
\begin{center}
\includegraphics[scale=0.65]{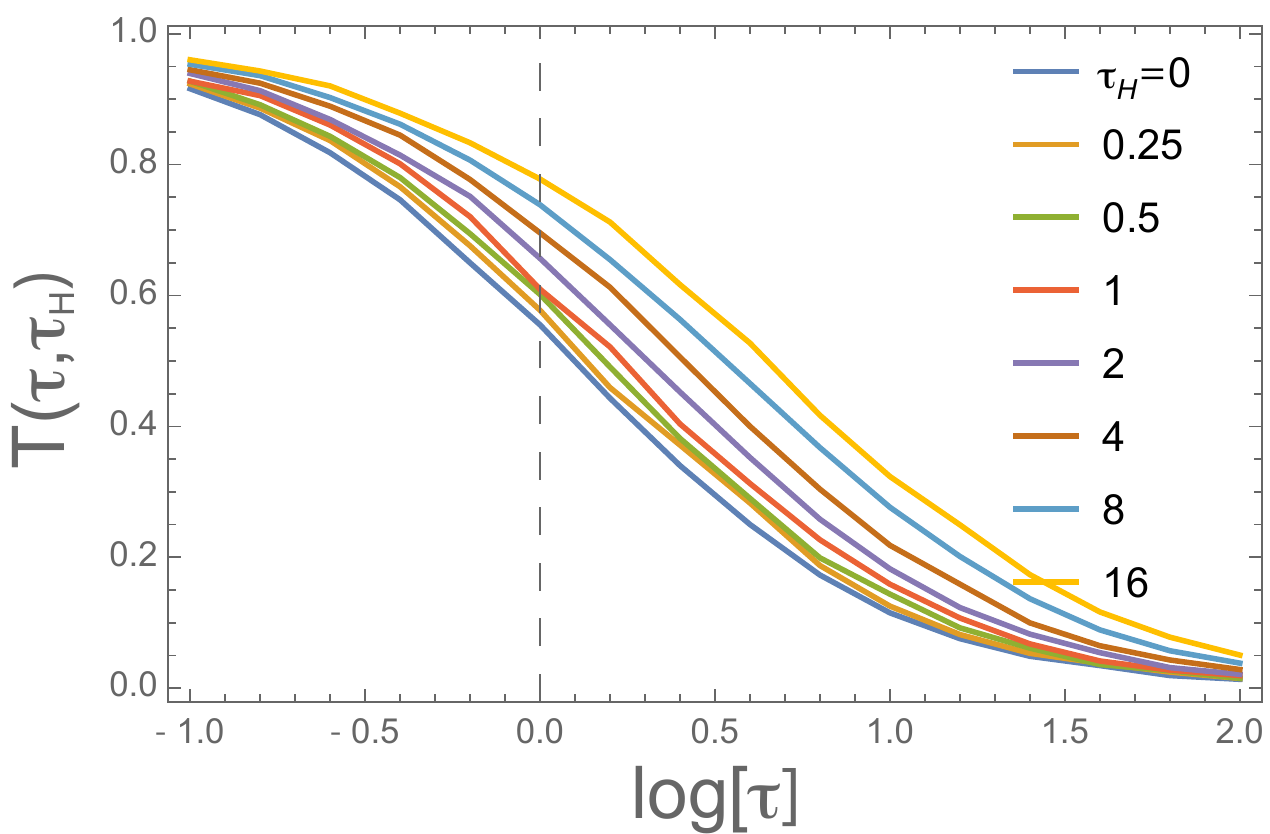}
\includegraphics[scale=0.34]{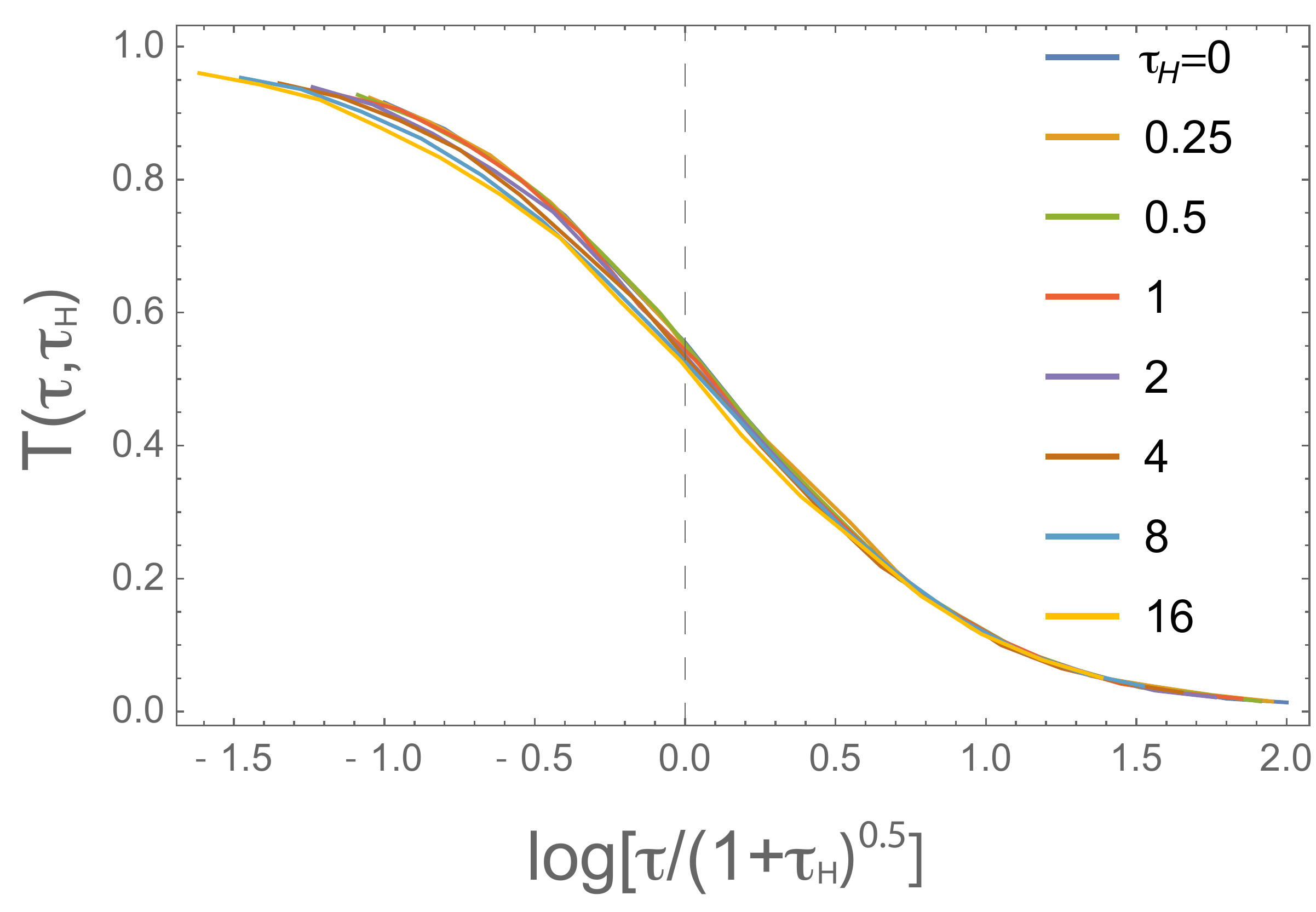}
\caption{
{\em Top:}
Results of Monte Carlo simulations for transmission $T$ through a porous scattering layer with extinction set by eqn.\ (\ref{eq;Etth}), plotted vs.\ mean optical depth $\tau$ for various porosity thicknesses $\tau_{\rm \hturb}$.
{\em Bottom:}
Same as top panel, but with the ordinate now the optical depth $\tau$ scaled by $\sqrt{1+\tau_{\rm \hturb}}$. The collapse of all cases to a nearly degenerate curve shows that porous transmission follows this scaling.
}
\label{fig:fig5}
\end{center}
\end{figure}

To illustrate explicitly this effect of porosity on transport through an optically thick medium, let us next consider the simple example of a planar layer with optical thickness $\tau$. 

For a pure-absorption opacity in a smooth, unclumped medium, the transmission of a vertical beam of radiation scales as $T(\tau) = e^{-\tau}$.
The above results now show that in medium with porosity thickness $\tau_{\rm \hturb}$, this pure-absorption transmission increases to $T(\tau,\tau_{\rm \hturb} ) = {\hat E} (\tau,\tau_{\rm \hturb})$.

In comparison, for the case of a pure-{\em scattering} layer, the transmission in a smooth medium now scales roughly as $T(\tau) \approx 1/(1+\tau)$, with thus much greater flux throughput than the exponential attenuation of a pure-absorption opacity.

To characterize the effect of porosity on such a scattering layer, we have carried out a simple Monte Carlo calculation using the porosity-modified form (\ref{eq:Ehatdef})
for the probability to transit a mean optical depth $\tau$ between each absorption/reemission.
At the bottom of a layer with mean vertical optical thickness $\tau$ and porosity thickness $\tau_{\rm \hturb}$, photons are introduced uniformly over the upward hemisphere, i.e.\ with some initial vertical direction cosine $0 \le \mu_o \le 1$.
Following each subsequent absorption, the photons are assumed to be scattered (or reemitted) {\em isotropically} into a new direction cosine $-1 \le \mu \le 1$.
Between each such scattering, the cumulative probability to move through a change in mean vertical optical depth $\Delta \tau$ is now given by
\beq
C(\Delta \tau/\mu,\tau_{\rm \hturb} )  = \int_0^{|\Delta \tau/\mu |} {\hat E} (t,\tau_{\rm \hturb}) \, dt
\, .
\label{eq:Cdef}
\eeq
For any given porosity depth $\tau_{\rm \hturb}$, we tabulate this function via numerical integration of (\ref{eq:Cdef}).
The Monte Carlo transport then proceeds as usual, by selecting uniform random numbers $0  \le x \le 1$, then solving for $\Delta \tau$ through inversion of the tabulation for $C(\Delta \tau/\mu, \tau_{\rm \hturb})=x$.
Each introduced photon is followed until it escapes either through the bottom or top of the layer, with the fraction escaping through the top providing then the net transmission $T$.

For the various labelled cases for the turbulent porosity depth $\tau_{\rm \hturb}$, the top panel of figure \ref{fig:fig5} plots this transmission vs.\ the layer's mean optical thickness $\tau$.
Note that higher values of $\tau_{\rm \hturb}$ lead to greater  transmission, implying an effective porosity reduction in the layer's optical thickness.

Indeed, the bottom panel of figure \ref{fig:fig5} shows that if one simply reduces this optical thickness by replacing $\tau \rightarrow \tau/\sqrt{1+\tau_{\rm \hturb}}$, the variation in transmission follows a nearly identical scaling for all the porosity models.
This provides a concrete example of how the {\em turbulent porosity effectively reduces the opacity} by a factor $1/\sqrt{1+\tau_{\rm \hturb}}$.

\section{Discussion and Future Outlook}

\subsection{Weakness of LDI porosity for bound-free continuum absorption of wind-broadened X-ray line emission}
\label{sec:weakpor}

In addition to generating clumping, the non-monotonic velocity resulting from the LDI leads to embedded wind shocks, with strong enough velocity jumps to produce soft 
($\gtrsim 0.5$\,keV) X-rays
\citep{Owocki08,Feldmeier97,Cohen14}.
This provides a natural explanation for the wind-broadened X-ray emission lines of lithium-like to hydrogen-like heavy ion transitions (e.g.\ SXIII, Si XIV, Fe XXIV) seen in high resolution X-ray spectra from OB 
supergiants (e.g., $\zeta$ Puppis)
 observed by the Chandra and XMM X-ray observatories
\citep{Ignace02,Cohen06,Cohen10}.
Since the longer-wavelength, red-shifted side of the emission profile originates from the opposite, receding hemisphere of the wind, it should suffer greater attenuation from bound-free 
(b-f) absorption by  relatively cool, inter-shock wind material.  As such, the X-ray emission profile from a dense wind is expected to exhibit a marked asymmetry to the short-wavelength, blue-shifted side.
Moreover,
since such b-f absorption scales linearly with wind density, this can provide a clumping-insensitive diagnostic of the wind mass loss rate
\citep{Cohen10}.
Indeed, in the several O-types with spectrally resolved wind-broadened X-ray emission lines, the relative modest level of observed asymmetry seems to require a mass loss rate that is reduced by a factor $\sim 3-6$ below what is commonly inferred from density-squared diagnostics, like radio or Balmer line-emission using smooth-wind 
models \citep[see][and references therein]{Cohen14b}.

A  caveat in these inferences has been the claim by some authors 
\citep[e.g.,][]{Feldmeier03,Oskinova04,Oskinova07}  that ``macro-clumping'' effects -- an alternative term for the porosity effects that obtain when clumps become optically thick -- could reduce the effectiveness of b-f absorption, and so explain the relatively symmetric X-ray profiles without such a marked reduction in mass loss rate.
 This could be the case for unspecified large-scale wind structures, e.g., those associated with discrete absorption components (DACs) \citep[see, e.g.,][and references therein]{Kaper99}, and/or arising from large-scale surface spots \citep{Ram17}. 
 
 But the LDI simulation results analyzed here strongly support previous arguments 
\citep{Owocki06,Sundqvist12b}
that clumping arising directly from the LDI itself has too small a scale to lead to significant macro-clumping or porosity effects\footnote{Note here that because of the high-mass loss rates from OB supergiant winds, the cooling length behind LDI-generated shocks is very small  \citep[see e.g.,][]{Owocki13}, implying that the hot, X-ray emitting gas occupies only a small fraction of the wind mass and volume. As such, even our strictly isothermal simulations should still provide a good description of the density structure in the cooled material that is responsible for b-f absorption of X-rays from small (and difficult to resolve) intervals of hot gas associated with embedded wind shocks.
Of course, such isothermal simulations do  {\em not} provide any direct model of the X-ray {\em emission} from such shock-heated gas.}.
Specifically, the above analysis indicates that LDI-generated structure has a turbulent porosity length, $\hturb \approx  \Delta X/4 = R_\ast/40 $; 
for winds with characteristic total continuum optical depth $\tau_\ast \equiv \kappa {\dot M}/4 \pi v_\infty R_\ast$, the associated porosity thickness would be thus be of order $\tau_{\rm \hturb} \approx \tau_\ast \hturb/R_\ast \approx  \tau_\ast/40$. 

Although the opacity from such b-f continuum X-ray absorption has a broad wavelength dependence, with sharp changes across ionization edges, the characteristic global optical depth $\tau_\ast$ is still only of order a few at all X-ray wavelengths \citep{Cohen14,Carneiro16}.
\citep[In particular, see, e.g., figure 10 of][for the case of the OB supergiant $\zeta$ Puppis]{Cohen10}.
As such, the much smaller porosity length $H \approx 0.025 R_\ast$ found here would be very optically thin ($\tau_{\rm \hturb} \ll 1$) for this full range of wavelengths, even at b-f edges. 
This implies LDI-generated porosity should give a negligible modification of X-ray transport in such OB-star winds\footnote{For future extensions to 
denser WR winds with  greater X-ray optical depths, the wavelength dependence of opacity can be readily accounted for through a ``multi-grey'' approach with separation into independent wavelength  bins.}.

In contrast, for the very optically thick UV resonance scattering lines, the strong LDI-generated clumping can lead to a {\em porosity in velocity space}  -- sometimes dubbed ``vorosity" \citep{Owocki08} -- that significantly reduces the saturation of the blueward absorption troughs of their observed P-Cygni line profiles.
As originally suggested by 
\citet{Oskinova07}, and extensively modeled by 
\citet{Sundqvist10,Sundqvist14}, this can provide a natural explanation for the unsaturated form of PV lines observed from many OB stars by the FUSE satellite mission 
\citep{Fullerton06}.

OB stars also show evidence for larger-scale surface variability associated with pulsations \citep{Simon-Diaz14,Simon-Diaz17} 
 and/or bright spots \citep{Ram17};
 these may play a key role in seeding co-rotating interaction regions (CIRs) \citep{Mullan86} or other large-scale wind structures associated with DACs and other variations in UV wind lines \citep{Cranmer96,David-Uraz17}.
In initial 2D simulations including both LDI and spot-induced CIRs \citep{Owocki99c}, the survival or disruption of such large-scale structure against the strong, small-scale LDI depended on numerical details, and so this remains an open issue.
To focus on the self-excited development of the small-scale LDI, the 2D simulations by \citet{Sundqvist17} did not include any external perturbations, but other 1D simulations 
\citep{Sundqvist13} show that these, along with other effects (e.g., limb darkening) that alter the transonic solution topology \citep{Sundqvist15}, can lead to development of extensive wind structure even close to the wind base, as is inferred from observations.
Thus while the analysis of these simulations here makes clear that LDI porosity should be at too small a scale to affect bound-free absorption of X-rays, the role of such larger-scale, surface-induced  wind structure for  both observed X-ray variability \citep{Oskinova14} and porosity reduction in their bound-free absorption remains an open issue.
This needs further exploration through multi-dimensional dynamical simulations, e.g. that account for both small-scale LDI and larger-scale, spot-induced CIRs using a less-restrictive formalism than developed in \citet{Owocki99c}.

\subsection{Potential dynamical role of turbulent porosity in optically thick outflows}

In addition to the potential effect on observational diagnostics, porosity can have dynamical effects in initiating and moderating radiatively driven wind outflows.
For example, 
\citet{Owocki04}
developed a model for super-Eddington winds in which the radiative driving is moderated by porosity effects under the assumption that the flow structure could be modeled as an exponentially truncated power-law in porosity length.
In the deeper atmosphere layers, the high optical thickness of the porosity length reduces the effective opacity and associated radiative acceleration to be below the local gravity, allowing a bound hydrostatic layer even when the gravitationally scaled smooth-medium acceleration $\Gamma = g_{\rm rad}/g = \kappa L_\ast/4 \pi GMc > 1$.
But for the lower-density upper layers, this porosity effect weakens, leading to a place where the scaled effective acceleration $\Gamma_{\rm eff} = 1$, and so to sonic-point initiation of a wind outflow.  By solving for the density $\rho_s$ at this sonic radius $r_s \approx R_\ast$ where the flow is at the sound speed $v_s$, one can derive an analytic scaling law for the mass-loss rate, ${\dot M} = 4 \pi r_s^2 \rho_s v_s$.
Thus another potential application of the turbulent porosity scalings derived here would be for such porosity-moderated  initiation of super-Eddington wind from a star with a continuum radiative acceleration that can overcome gravity.

A related issue regards the role of the iron opacity bump at temperatures $T \sim 150,000$\,K in initiating the strong, line-driven stellar winds of Wolf-Rayet stars 
\citep{Nugis02}, and in possibly inducing a density inversion and associated inflation of the stellar envelope
\citep[see][and references therein]{Sanyal17}.
Recent 3D radiation-hydrodynamics simulations by 
\citet{Jiang15,Jiang17} show that density inversion that can form near this opacity bump tends to break up into complex, 3D turbulent structure with super-sonic flows and associated density structure. 
Particularly for the MHD models \citep{Jiang17} with more transonic velocity variations and thus greater density fluctuations, this leads to porosity reduction of the opacity and associated radiative acceleration, with then a feedback on the density inversion and stratification. 
It would thus be interesting to 
apply the turbulent porosity formalism here  to such 3D radiation-hydrodynamical simulations, and compare the derived porosity scalings with direct computations of the transport and its effect on radiative acceleration.

Finally, it would also be interesting to explore the potential relevance of the general formalism developed here for the broader study of porosity effects in astrophysical media with complex structure in density and opacity, e.g., for the interstellar medium.

\section*{Acknowledgments}
This work was supported in part by SAO Chandra grant
TM3-14001A and NASA grant NNX15AM96G awarded to the University of Delaware,
and in part by a visiting professor scholarship ZKD1332-00-D01 for SPO from KU Leuven and its Institute voor
Sterrenkunde (IvS).
SPO also acknowledges sabbatical leave support from the University of Delaware.
We thank J. Puls for helpful comments on statistical distributions and related issues.

\bibliographystyle{mn2e}
\bibliography{OwockiS}

\end{document}